\documentstyle[prl,aps,multicol,epsf,rotate]{revtex}
\def\ffrac#1#2{\textstyle{#1\over#2}\displaystyle}
\begin{document}
%
\title{Crossing Formulae for Critical Percolation in an Annulus}
\author{John Cardy}
\address{University of Oxford, Department of Physics -- Theoretical
         Physics\\
         1 Keble Road, Oxford OX1 3NP, U.K. \\
         and All Souls College, Oxford.}
%
%
\maketitle
\begin{abstract}
An exact formula is given for the probability that there exists a spanning
cluster between opposite boundaries of an annulus, in the scaling limit
of critical percolation. The entire distribution function for the number
of distinct spanning clusters is also derived. These results are found 
using Coulomb gas methods. Their forms are compared with the expectations of
conformal field theory.
\end{abstract}
%
%
\vspace{1cm}
Since Langlands et al.\cite{Lang}
conjectured on the basis of numerical evidence
that crossing probabilities between two 
non-overlapping segments of the boundary of a simply connected region
should be conformally invariant, there has been intense interest in the
scaling limit of two-dimensional percolation\cite{Aizenman}.
In \cite{Cardy1} it was shown that this invariance was implicit in
ideas of conformal field theory, which in
addition yielded an explicit formula. Further exact formulas were
conjectured, by Watts\cite{Watts}
for the probability of a simultaneous left-right
and up-down crossing, and by Pinson\cite{Pinson}
for various crossing probabilities on the torus. The latter work used
so-called Coulomb gas methods\cite{Nienhuis}, which had been developed for more
general two-dimensional critical systems, in parallel with those of
conformal field theory. In \cite{Cardy2} results were conjectured for
the asymptotic behaviour of the probabilities that at least $N_c$
distinct clusters cross either a rectangle or an annulus,
using earlier conjectures of Saleur and Duplantier\cite{DupSal}.
In \cite{Cardy3},
among other results, a prediction was given for the mean number of
crossing clusters in the opposite limit, when this number is large.

Meanwhile, starting from another approach, Schramm\cite{Schramm}
conjectured that the
scaling limit of percolation hulls is generated by stochastic Loewner
evolution (SLE$_6$). From this follow many results\cite{LSW}, including the
original crossing formula and the exponents in \cite{Cardy2}. Finally 
Smirnov\cite{Smirnov} proved the original crossing formula for site
percolation on the triangular lattice,
and hence the validity of the SLE$_6$ approach to percolation\cite{SW}.

In this paper we refine the results of \cite{Cardy2} for the annulus,
presenting results for a general value of the modulus.
Consider a critical percolation problem in a non-simply connected region
of the plane with the topology of an annulus. The boundaries are assumed
to be suitably smooth. The interior of this region may be conformally
mapped into the interior of a circular annulus $R_1\leq|z|\leq R_2$,
with modulus $\tilde q\equiv(R_1/R_2)$, or into the rectangle $(0\leq x<\ell,
0<y<L)$ with the edges at $x=0$ and $x=\ell$ identified, and
$\tilde q=e^{-2\pi L/\ell}$.   
A crossing (or spanning)
cluster is one which contains a path connecting the opposite
boundaries.
Let $P(N_c)$ be the probability that there are exactly $N_c$ non-overlapping
such clusters. When $N_c=1$ it is possible for the cluster also
to wrap around the $x$-cycle on the annulus. By convention, we {\bf do not}
count such clusters as spanning.

\subsection*{Results.}
The crossing probability is
\begin{equation}
\label{x1}
\sum_{N_c=1}^\infty P(N_c)=
\sqrt3\,
{\sum_{r\in{\bf Z}}\left({\tilde q}^{12r^2+4r+\frac14}
-{\tilde q}^{12r^2+8r+\frac54}\right)\over
\sum_{r\in{\bf Z}}\left({\tilde q}^{12r^2+2r}
-{\tilde q}^{12r^2+10r+2}\right)}
\end{equation}
Furthermore we have an explicit expression for $P(N_c)$ for $N_c\geq1$:
\begin{equation}
\label{PN}
P(N_c)={3^{N_c-\frac12}\over2^{2N_c-1}}\prod_{n=1}^\infty(1-{\tilde
q}^{2n})^{-1}\sum_{s=0}^\infty A_s(N_c)\,{\tilde q}^{\frac{(N_c+s)^2}3-
\frac1{12}}
\end{equation}
where
\begin{equation}
\label{AN}
A_s(N_c)=(-1)^s\sum_{r=s}^{N_c+s}{r\choose s}{{2N_c+2s}\choose{2r}}
\quad.
\end{equation}
These results may be transformed into other expressions using
various theta-function identities. For example, in terms of the conjugate
modulus $q\equiv e^{-\pi\ell/L}$, we find
\begin{equation}
\label{x2}
\sum_{N_c=1}^\infty P(N_c)=
{\sum_{r\in{\bf Z}}\left(q^{6r^2+r}+q^{6r^2+5r+1}-2q^{6r^2+3r+\frac13}
\right)\over
\sum_{r\in{\bf Z}}\left(q^{6r^2+r}-q^{6r^2+5r+1}\right)}\quad.
\end{equation}
Note that for $L/\ell>\frac12$ only a few terms need be kept in
(\ref{x1},\ref{PN}) for great accuracy, while for $L/\ell<\frac12$ the same is
true of (\ref{x2}). 

Both the numerator and denominator of (\ref{x1}) and (\ref{x2}) 
are specialisations of
Jacobi theta functions, and hence may be written as infinite products.
In terms of Dedekind's eta function 
$\eta(\tau)\equiv {\tilde q}^{1/24}\prod_{n=1}^\infty(1-{\tilde q}^n)$,
with $\tilde q\equiv e^{2\pi i\tau}$, we find
\begin{equation}
\label{x3}
\sum_{N_c=1}^\infty P(N_c)=
\sqrt3\,{\eta(\tau)\,\eta(6\tau)^2\over\eta(3\tau)\,\eta(2\tau)^2}
={\eta(-1/\tau)\,\eta(-1/6\tau)^2\over\eta(-1/3\tau)\,\eta(-1/2\tau)^2}
\quad.
\end{equation}

\subsection*{Coulomb gas method.}
\begin{figure}
$$
\epsfxsize=4.5in
\epsfbox{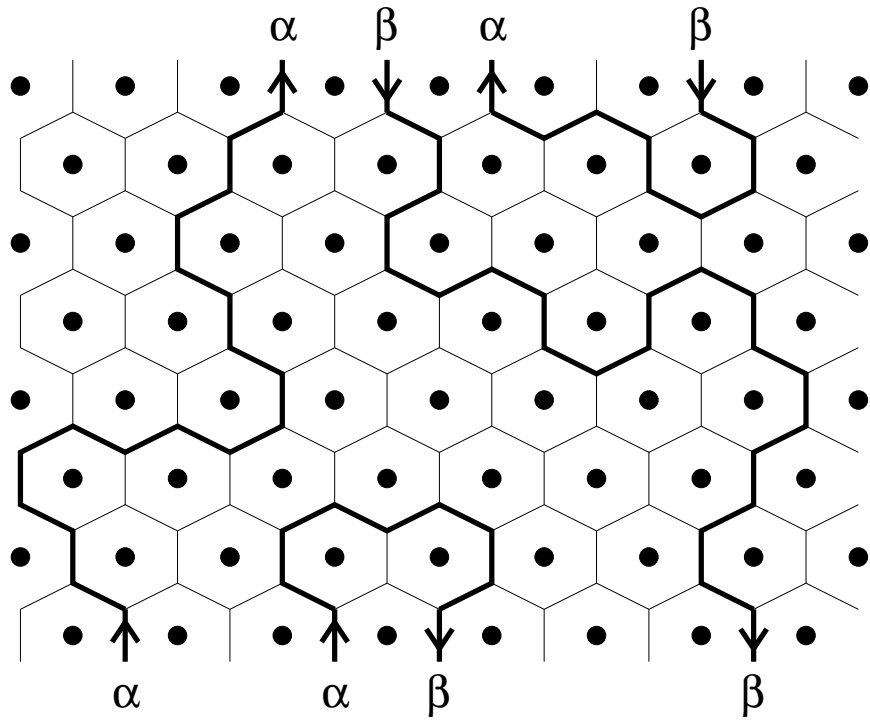}
$$
\caption{A triangular lattice with $\ell=8a$, $L=7(\sqrt3/2)a$.
The leftmost and rightmost columns are to be identified, so the lattice
has the topology of an annulus. Typical oriented spanning and
non-spanning open
hulls are shown, together with their boundary weights.}
\label{fig1}
\end{figure}

Although we shall later argue that these results are indeed conformally
invariant, it is simpler first to set the problem up in the periodic
rectangle defined above.
Consider a portion of a regular triangular lattice covering the rectangle,
oriented as shown in
Fig.~\ref{fig1}, so that, if the lattice spacing is $a$, there are 
$2(\ell/a)+1$ columns and $(2/\sqrt3)(L/a)+1$ rows of the lattice.
Impose periodic boundary conditions in the $x$-direction, so that the 
rightmost column is identified with the leftmost one. 
Consider a critical site percolation problem on this lattice, in which
sites are independently coloured red or blue with equal probability.
A (blue) cluster is a set of blue sites in which every site is connected to
every other by a path which traverses only blue sites. 
A spanning cluster is one which contains at least one site on
the edge $y=0$ and one site on the edge $y=L$.  For a
particular assignment of colours, let $N_c$ 
be the number of distinct non-overlapping
spanning clusters.
We are interested in the distribution $P(N_c)$ of the random
variable $N_c$ in the continuum limit $a\to0$, for fixed $L$ and $\ell$.
Scale invariance implies that it should depend only on $\ell/L$.

Instead of considering the clusters, we may equivalently consider the
configuration of hulls which separate them. A hull is a path on the dual
lattice (in this case a honeycomb lattice) which separates blue sites
from red sites. In our case, hulls can either form closed paths or be
open, each end terminating at an edge. Open hulls which have ends terminating
on different edges are called spanning hulls. The number of such
spanning hulls is $n_c=2N_c$.
An allowable configuration of hulls is one in which each dual site is
connected to either 0 or 2 neighbouring dual sites, except for
the edge sites, which may be connected to either 0 or 1 neighbouring
site. In addition, the number of spanning hulls must be even. The correct
weights are achieved by weighting all allowable hull configurations equally.

A related model is the O$(1)$, or Ising, model
on the dual lattice, at zero temperature. In this model Ising spins
$s({\bf r})=\pm1$ reside at each site ${\bf r}$ of the honeycomb lattice, and the
partition function is 
\begin{equation}
Z_{{\rm O}(1)}=\prod_{\bf r}
\sum_{s({\bf r})=\pm1}\prod_{({\bf r},{\bf r}')}
\ffrac12\left(1+ts({\bf r})s({\bf r}')\right)
\end{equation}
where the latter product is over all nearest neighbour pairs 
$({\bf r},{\bf r}')$.
The `high-temperature'
expansion in powers of $t$, afterwards setting $t=1$, reproduces
exactly the hull configurations of the percolation model in the case
when the number of spanning hulls, 
denoted by $n_c$, is even and equal to $2N_c$,
but in the O$(1)$ model $n_c$ may also be odd. Evidently when $t=1$,
$Z_{{\rm O}(1)}=2$. Denoting by $p(n_c)$ the probability that there are exactly
$n_c$ spanning hulls, we have $P(N_c)=2p(2n_c)$.
The first factor of $2$ arises because to each allowable configuration of   
hulls there correspond two assignments of colours.
We can construct the generating function 
by weighting each spanning hull by a factor $u$. 
Denoting the corresponding partition function by $Z(u)$ we therefore
have
\begin{eqnarray}
\sum_{n_c=0}^\infty p(n_c)u^{n_c}&=&\ffrac12Z(u)\\
\sum_{N_c=0}^\infty P(N_c)u^{2N_c}&=&\ffrac12\big(Z(u)+Z(-u)\big)
\end{eqnarray}

Each hull may assigned a random orientation, so that
to each configuration of $H$ hulls correspond $2^H$ configurations
of oriented hulls. The weights for each orientation should be chosen
so that they sum to unity (resp.~$u$) for each (spanning) hull.
For closed hulls, it is conventional\cite{Nienhuis}
to assign a `weight' $e^{\pm
i\chi}$ to each dual site at which an oriented hull turns through an
angle of $\pm\pi/3$, where $\chi=\pi/18$ is chosen so that the total weight for
a closed hull, on summing over its orientations, is $e^{6i\chi}+
e^{-6i\chi}=1$. However, this does not correctly account for closed
hulls which wrap around the $x$-cycle of the annulus, 
which would have weight $1+1=2$
according to this scheme. 
Such configurations can only occur when $n_c=0$. Thus, for the time
being, we assume that $n_c\geq1$. The case $n_c=0$ may be
inferred afterwards using the overall normalisation of the partition
function.

For oriented hulls which terminate at an edge, let us 
assign the same weights as above for internal turnings, and in addition weights
$\alpha$ or $\beta$ to their
extreme segments, as shown in
Fig.~\ref{fig1}. By choosing 
\begin{equation}
\label{weights}
\alpha=\left({\cos6\chi\over\cos3\chi}\right)^{1/2}
\,e^{3i\chi'/2}\quad{\rm and\ }\quad
\beta=\left({\cos6\chi\over\cos3\chi}\right)^{1/2}
\,e^{-3i\chi'/2}\quad,
\end{equation}
hulls which begin and end on the same edge are counted with a weight
$(\cos6\chi/\cos3\chi)(e^{3i\chi}+e^{-3i\chi})=1$, 
as required, while spanning hulls carry a weight
$(\cos6\chi/\cos3\chi)(e^{3i\chi'}+e^{-3i\chi'})$,
so that we should identify
\begin{equation}
\label{u}
u\equiv \cos3\chi'/\cos3\chi\quad.
\end{equation}

The factor $(\cos6\chi/\cos3\chi)^{1/2}$ coming from (\ref{weights})
is raised to a power $E$ which is the total number of ends of
open hulls, whether they be spanning or not. 
An open end occurs
every time the neighbouring sites of the triangular lattice
are of opposite colours. Since these are independently distributed,
$E$ is a sum of $O(2\ell/a)$ independent\footnote{Almost independent,
since the sum along each edge must be even.} random variables, 
each taking the values
$0$ or $1$ with equal probability. In the continuum limit $a/\ell\to0$,
therefore,
$(\cos6\chi/\cos3\chi)^{E/2}\sim(\cos6\chi/\cos3\chi)^{\ell/2a}$, with
probability one. These contribute to the non-universal edge free energy,
but not to the universal dependence on $\ell/L$.

Let $Z(3\chi,3\chi')$ be the partition function of the loop gas with the
above phase factors but ignoring the factors of
$(\cos6\chi/\cos3\chi)^{1/2}$. Then
\begin{equation}
\label{gf1}
\sum_{n_c=1}^\infty p(n_c)u^{n_c}
=C_1\left(Z(\pi/6,\chi')-Z(\pi/6,\pi/2)\right)
\end{equation}
where $C_1$ is a non-universal number and the second term,
with $\cos3\chi'=0$, subtracts out the contribution with $n_c=0$ which
is incorrectly counted by the above scheme.

The configurations of the oriented loop gas are in 1-1 correspondence
with those of a height model on the original triangular lattice.
These heights $h(r)$ are conventionally chosen to be in $\pi{\bf Z}$,
and are defined by the conditions that $h=0$ at some fixed site, say
$(0,0)$, and that it increases (decreases) by $\pm\pi$ each time an
oriented hull segment is crossed. On the annulus, however, we must also
allow for possible jumps $\Delta h>|\pi|$ in $h$ across some path which spans
the annulus, say along $x=-\frac14a$. The factors $e^{\pm3i\chi'/2}$ then
accumulate to $e^{3i\chi'\Delta h/2\pi}$ on each edge.

So far, everything is finite and exact. 
In the conventional Coulomb gas method\cite{Nienhuis}, one
now assumes that in the continuum limit $(a/\ell,a/L)\to0$ we may
replace $h(r)$ by a real-valued field, with a gaussian measure
$\propto\exp\big(-(g/4\pi)\int(\partial h)^2dxdy\big)$. For the models
we are considering, $g$ is fixed to be $1-(6\chi/\pi)=\frac23$. 
We shall assume that the same is true on the annulus, except that we
must allow for a possible discontinuity around the $x$-cycle.
Thus we write
\begin{equation}
h(x,y)=(p\pi/\ell)x +\tilde h(x,y)
\end{equation}
where $p\in{\bf Z}$ and $\tilde h(x+\ell,y)=\tilde h(x,y)$.
Substituting in this decomposition,
\begin{equation}
\label{C2}
Z(3\chi,3\chi')=C_2{\cal Z}(\ell/L)\sum_{p\in{\bf Z}}e^{3i\chi'p}
e^{-(g/4\pi)(p\pi/\ell)^2\ell L}
\end{equation}
where ${\cal Z}\propto
\int{\cal D}\tilde h e^{-(g/4\pi)\int(\partial\tilde h)^2dxdy}$
is the universal part of the partition function of a free field on the
annulus, with Neumann boundary conditions, and with the constraint that
$\tilde h({\bf 0})=0$, which removes the zero mode.
The factor $C_2$ is non-universal, and reflects the contribution of the
short-distance degrees of freedom which are integrated out in the
coarse-graining assumed in adopting the gaussian measure. It is expected
to depend exponentially on the total area $(\ell L/a^2)$ and the perimeter
$(2\ell/a)$, but is not expected to have nontrivial dependence on
the modulus $\ell/L$. 

The $c=1$ partition function $\cal Z$ is well-known\cite{Itzyk}. Writing it
as ${\rm Tr}\,e^{-\ell\hat H_L}$, where $\hat H_L$ is the quantum
hamiltonian for a free field on circle of perimeter $L$, it is
$\prod_{n=1}^\infty\sum_{N=0}^\infty e^{-\ell E_{n,N}}$ where 
$E_{n,N}=(N+\frac12)(n\pi/L)$. The leading term as $\ell/L\to\infty$
comes from $N=0$ and is proportional to $\prod_{n=1}^\infty
e^{-(\pi\ell/2L)n}$. This must be regularised, however. Apart from a
cut-off dependent term which can be absorbed into $C_2$, it gives
$e^{-(\pi\ell/2L)\zeta(-1)}=q^{-\frac1{24}}$, where $q\equiv
e^{-\pi\ell/L}$. The terms with $N\geq1$
give $\prod_{n=1}^\infty(1-q^n)^{-1}$. The zero-mode 
$\tilde h=$ constant is suppressed in the functional integral over
$\tilde h$ since we set $\tilde h({\bf 0})=0$. However going from this
constraint to one on the $n=0$ mode introduces a jacobian
proportional to $(L/\ell)^{1/2}$ \cite{Itzyk}. Finally we have
\begin{equation}
\label{C3}
{\cal Z}=C_3(L/\ell)^{1/2}q^{-\frac1{24}}\prod_{n=1}^\infty(1-q^n)^{-1}
\quad.
\end{equation}

Eq.~(\ref{C2}) may now be transformed using the Poisson sum formula: 
\begin{eqnarray}
Z(3\chi,3\chi')&=&
\label{C4}
C_4\,{\cal Z}\,\sum_{r\in{\bf Z}}\int_{-\infty}^\infty dp\,
e^{2\pi ipr}\,e^{3i\chi'p}\,e^{-(\pi g/4)(L/\ell)p^2}\\
\label{C5}
&=&C_5\,q^{-\frac1{24}}\prod_{n=1}^\infty(1-q^n)^{-1}\sum_{r\in{\bf Z}}
e^{-((3\chi'+2\pi r)^2)/\pi g)(\ell/L)}
\end{eqnarray}
Note that the $(L/\ell)^{1/2}$ factors cancel.

Setting now $g=\frac23$ and subtracting the contributions with
$3\chi'=\frac\pi6$ and $3\chi'=\frac\pi2$, we arrive, after some
algebra, at the result for the $O(1)$ model
\begin{equation}
\label{O1}
\sum_{n_c=1}^\infty p(n_c) = C_5\,
{\sum_{r\in{\bf Z}}\left(q^{6r^2+r}-q^{6r^2+3r+\frac13}\right)\over
\prod_{n=1}^\infty(1-q^n)}
\end{equation}

Because our height model phase assignments incorrectly count loops which
wrap around the $x$-cycle, we cannot directly compute the contribution
with $n_c=0$ and therefore cannot fix $C_5$ by demanding that
$\sum_{n_c=0}^\infty p(n_c)=1$. 
However, since $\sum_{N_c=0}^\infty P(N_c)=2\sum_{N_c}p(2N_c)=1$,
it follows that $\sum_{n=0}^\infty p(2n+1)=\frac12$, and we can
compute this in terms of
$Z(u)-Z(-u)\propto Z(\pi/6,\pi/6)-Z(\pi/6,5\pi/6)$. This gives
\begin{equation}
\sum_{n_c\ {\rm odd}}p(n_c)=\ffrac12C_5
{\sum_{r\in{\bf Z}}\left(q^{6r^2+r}-q^{6r^2+5r+1}\right)\over
\prod_{n=1}^\infty(1-q^n)}=\ffrac12C_5\,
\end{equation}
where the last equality follows from
Euler's pentagonal number theorem\cite{Euler}. We conclude that $C_5=1$.
(\ref{O1}) is then our main result for the probability of a crossing in
the $O(1)$ model.

For percolation, we need to compute $2\sum_{N_c=1}^\infty p(2N_c)
\propto Z(\pi/6,\pi/6)+Z(\pi/6,5\pi/6)-2Z(\pi/6,\pi/2)$. This gives the
main result (\ref{x2}).

With the knowledge that $C_5=1$, we may now transform these results back
into series in $\tilde q\equiv e^{-2\pi L/\ell}=e^{2\pi i\tau}$.
Using the identity
$\eta(\tau)=(-i\tau)^{-\frac12}\eta(-1/\tau)$,
and the Poisson sum formula, we find
\begin{equation}
\label{xxx}
Z(\pi/6,\chi')=\frac1{2\sqrt3}\prod_{n=1}^\infty(1-{\tilde q}^{2n})^{-1}
\sum_{p\in{\bf Z}}e^{3i\chi'p}\,{\tilde q}^{(p^2-1)/12}
\end{equation}
Thus for the crossing probability in percolation we have
\begin{equation}
\sum_{N_c=1}^\infty P(N_c)=
\frac1{2\sqrt3}\prod_{n=1}^\infty(1-{\tilde q}^{2n})^{-1}
\sum_{p\in{\bf Z}}\left(\cos\ffrac{\pi p}6+\cos\ffrac{5\pi p}6
-2\cos\ffrac{\pi p}2\right)\,{\tilde q}^{(p^2-1)/12}
\end{equation}
The expression in parentheses takes the value $3$ if $p=\pm2\,({\rm mod\ }12)$,
the value $-3$ if $p=\pm4\,({\rm mod\ }12)$, and vanishes otherwise. This
leads to the first form (\ref{x1}) of our main result.
The numerator in this expression may also be written as\cite{theta}
\begin{eqnarray}
\sum_{n\in{\bf Z}}(-1)^n{\tilde q}^{3n^2+2n+\frac14}
&=&{\tilde q}^{\frac14}\vartheta_4(2\tau|6\tau)\\
&=&{\tilde q}^{\frac14}
\prod_{n=1}^\infty\left((1-{\tilde q}^{6n})(1-{\tilde q}^{6n-1})
(1-{\tilde q}^{6n-5})\right)\quad,
\end{eqnarray}
which, after a few more manipulations, gives (\ref{x3}).

In order to find an explicit formula for $p(n_c)$, we should solve
(\ref{u}) for $e^{3i\chi'}$ in terms of $u$, which gives 
\begin{equation}
e^{3i\chi'}=e^{i\pi/2}\left((1-(\sqrt3u/2)^2)^{1/2}-e^{-i\pi/2}
(\sqrt3u/2)\right)\quad.
\end{equation}
Substituting this into (\ref{xxx}), 
expanding in powers of $u$, and identifying the coefficient of
$u^{2N_c}$, then leads to the result in (\ref{PN},\ref{AN}).

A further check on our results comes from differentiating (\ref{gf1})
with respect to $u$ at $u=0$ to obtain the mean number of crossing
clusters. In the limit $\ell\gg L$ we find 
$E[N_c]\sim (\sqrt3/4)(\ell/L)$, in agreement with \cite{Cardy3}, and
with a rigorous result of Smirnov\cite{Smirnov} for the triangular
lattice. 

If instead of the periodic rectangle we have a more general annular
region, in order that
spanning and non-spanning hulls be counted with their correct weights
$\alpha_i$ and $\beta_i$ in (\ref{weights}) must be
modified by factors $e^{\pm i\theta/6}$, where $\theta$ is the (signed)
angle which the tangent vector at the boundary makes with the $x$-axis.
However, since the boundaries form simple closed curves, these extra
factors accumulate to unity on each edge.
The calculation then proceeds as
before, yielding a conformally invariant result.\footnote{
Under a scale transformation ${\bf r}\to\lambda{\bf r}$, a partition
function $Z$ behaves in general as $\lambda^{c\chi/6}$, where $c$ is the
conformal anomaly number, and, for a smooth boundary, $\chi$ is the
Euler number\cite{CardyPeschel}.
The latter vanishes for the annulus. If there are points
on the boundary where it is not differentiable, however, there may be
additional contributions\cite{CardyPeschel}. 
In our case, these cancel between the $c=1$
partition function $\cal Z$ and the Coulomb energy of the charges which
accumulate at these singularities. This cancellation is connected with
the fact that the overall conformal field theory has $c=0$.}

\subsection*{Relation with conformal field theory.}
The crossing probability $\sum_{n_c=1}^\infty
p(n_c)$ in the O$(1)$ model may be expressed as a difference 
$Z_{++}-Z_{+-}$ of partition functions in the $n\to1$ limit of the
O$(n)$ model, where $Z_{+-}$ denotes the partition function with the
spins fixed in different directions on opposite edges, 
and $Z_{++}$ with them fixed in the same direction. Evidently 
$Z_{++}=1$, so that, from (\ref{O1})
\begin{equation}
\label{Zab}
Z_{+-}= 
{\sum_{r\in{\bf Z}}\left(q^{6r^2+3r+\frac13}-q^{6r^2+5r+1}\right)
\over\prod_{n=1}^\infty(1-q^n)}
\end{equation}
According to general boundary conformal field theory (BCFT)\cite{Cardybcft}, 
any partition
function like this should be expressible in the form
$\sum_hd_hq^h$, where $h$ runs over all boundary scaling dimensions
and $d_h$ is a degeneracy factor. For unitary conformal theories
this must be a non-negative integer, but this need not be true
here. From (\ref{Zab}) we identify the smallest value
of $h$ to be $\frac13$: this is identified in BCFT\cite{Cardybcft}
as the scaling dimension of
the `boundary condition changing operator' $\phi_{+|-}$. This is
consistent with the analogous result for percolation: see
\cite{Cardy1}. From (\ref{Zab}) we see there is also an operator
with $h=1$. This we tentatively identify as introducing a hull which
wraps around the $x$-cycle in the O$(1)$ model, but does not touch
either edge. This should carry an
O$(n)$ index $c$ which is not equal to either $+$ or $-$, otherwise 
it could be absorbed at the edges. There are $n-2=-1$ possibilities for
$c$, which accounts for the fact that this state occurs with degeneracy
$(-1)$ in (\ref{Zab}). 

The powers $(4p^2-1)/12$ in (\ref{PN}) are the well-known
bulk multi-hull dimensions for percolation\cite{DupSal}. 
In accordance with general ideas of BCFT\cite{Cardybcft}, (\ref{Zab})
may be written as a linear combination of Virasoro
characters $\chi_h(q)$ of irreducible representations of highest weight
$h$, and, equivalently, as a combination of characters $\chi_h({\tilde
q}^2)$, related by a modular transformation. This affords an
interesting example of how BCFT works in a non-minimal theory.
The details will be described elsewhere\cite{JCinprep}.
\subsection*{Comparison with numerical work.}
Shchur\cite{Shchur} has estimated
$P(N_c)$ for $N_c=2,3$ and $L=\ell$. For this value of $\tilde
q=e^{-2\pi}$ it is sufficient to keep only the first term in (\ref{PN}),
$P(N_c)\approx 3^{N_c-\frac12}{\tilde q}^{(4N_c^2-1)/12}$,
which gives
\begin{eqnarray}
P(2)&=& 2.02\ldots\times 10^{-3}\quad({\rm exact}),\quad
        2.0(4)\times 10^{-3}\quad({\rm measured});\\
P(3)&=& 1.71\ldots\times 10^{-7}\quad({\rm exact}),\quad
        1.4(5)\times 10^{-7}\quad({\rm measured}).
\end{eqnarray}
Our exact predictions fall within the (admittedly rather large) error
bars.  
\subsection*{Summary.}
We have given explicit results for the probability that $N_c$ critical
percolation clusters cross an annulus. From the point of view of
conformal field theory, these results are different from the original
crossing formula\cite{Cardy1} in that they involve partition functions
rather than correlation functions of boundary operators. 
The exponents appearing in (\ref{PN}) have already been derived
in the limit of large modulus using a radial version of
SLE\cite{LSW,SW}, and it would be very interesting to use these methods
to verify the more detailed results of the present paper.

\subsubsection*{Acknowledgements} This work was begun while the author
visited the KTH and the Mittag-Leffler Institute, Stockholm. He
thanks S.~Smirnov and 
these institutes for their hospitality, and L.~Carleson for
asking the question. He also thanks R.~Ziff for comments on an earlier
version of this paper. This work was also supported in part by
EPSRC Grant GR/J78327.

\end{document}